\documentclass[aps,prl,twocolumn,superscriptaddress]{revtex4-1}
\usepackage{graphicx}
\usepackage{subfig}
\usepackage{floatrow}
\usepackage{amsmath}
\usepackage{hyperref}
\usepackage{floatrow}
\usepackage{siunitx}
\usepackage {ulem}

\usepackage{ragged2e}
\usepackage{dcolumn}
\usepackage{bm}

\begin{document}

\preprint{APS/123-QED}

\title{Propagation, dissipation and breakdown in quantum anomalous Hall edge states probed by microwave edge plasmons}
%

\author{Torsten Röper}
\author{Hugo Thomas}
\author{Daniel Rosenbach}
\author{Anjana Uday}
\author{Gertjan Lippertz}
 \affiliation{II. Physikalisches Institut, Universit\"at zu K\"oln, Z\"ulpicher Str. 77, D-50937 K\"oln, Germany}
\author{Anne Denis}
\author{Pascal Morfin}
\affiliation{Laboratoire de Physique de l'Ecole Normale Supérieure, ENS, Université PSL,
CNRS, Sorbonne Université, Université de Paris, F-75005 Paris, France}
\author{Alexey A. Taskin}
\author{Yoichi Ando}
\author{Erwann Bocquillon}%
 \email{bocquillon@ph2.uni-koeln.de}
 \affiliation{II. Physikalisches Institut, Universit\"at zu K\"oln, Z\"ulpicher Str. 77, D-50937 K\"oln, Germany}

\date{\today}


\begin{abstract}
The quantum anomalous Hall (QAH) effect, with its single chiral, topologically protected edge state, offers a platform for flying Majorana states as well as non-reciprocal microwave devices. While recent research showed the non-reciprocity of edge plasmons in Cr-doped $\mathrm{(Bi_xSb_\text{1-x})_2Te_3}$, the understanding of their dissipation remains incomplete. Our study explores edge plasmon dissipation in V-doped $\mathrm{(Bi_xSb_\text{1-x})_2Te_3}$ films, analyzing microwave transmission across various conditions. We identify interactions with charge puddles as a primary source of dissipation, providing insights critical for developing improved QAH-based technologies.

\end{abstract}

\maketitle


\paragraph{\label{sec:Intro} Introduction --} 
    Magnetic topological insulators exhibit the quantum anomalous Hall effect \cite{chang_experimental_2013,chang_high-precision_2015,Yu2010},  namely the occurrence of a single chiral edge state, carrying current in a direction determined by the material's magnetization direction, while the bulk remains insulating. Despite magnetic and electronic disorder, quantized dissipationless edge transport is observed in absence of a magnetic field at very low temperatures. This robust topological phase of matter opens interesting perspectives for quantum metrology \cite{Bestwick-2015,gotz2018,Fox-2018,fijalkowski2024}, and non-reciprocal microwave components \cite{viola_hall_2014, mahoney_zero-field_2017}. When in proximity with a superconductor \cite{uday2023}, these edge states allow for the creation of flying non-abelian anyons, which could be braided as they propagate \cite{beenakker_deterministic_2019,beenakker_electrical_2019}. A pending issue is in particular the origin of the breakdown of the QAH regime at larger temperatures or currents, attributed to either heating effects \cite{Fox-2018,ferguson2023} as in the quantum Hall effect \cite{komiyama2000} or to a field-induced percolation of charge puddles created by disordered fluctuations of the Fermi level \cite{lippertz-2022}. Prospective applications require a good understanding and control of the high-frequency propagation of plasmons in the edge state. To date, two studies have focused on narrow-band resonant geometries in Cr-doped $\mathrm{(Bi_x Sb_\text{1-x})_2 Te_3}$  disks, and reported a circulation effect in this material, with a first investigation of losses \cite{mahoney_zero-field_2017,martinez_edge_2023}. The screening and dissipation due to charge puddles deserve particular attention.
    Here, we report on a systematic and broadband study of plasmons in the QAH insulator V-doped $\mathrm{(Bi_x Sb_\text{1-x})_2 Te_3}$ (V-BST), a material which offers better performance and a more stable QAH state in zero magnetic field owing to its larger coercive field ($H_c\simeq\SI{0.9}{\tesla}$) compared to Cr-doped BST ($H_c\simeq\SI{0.2}{\tesla}$) \cite{chang_experimental_2013,Bestwick-2015,Fox-2018}. Using a microwave Hall bar geometry with capacitively coupled input and readout ports, we carry out broadband measurements from 0.3 to \SI{4}{\giga\hertz}. We resolve the phase velocity and the dissipation of plasmons from the phase and amplitude of the transmission coefficient respectively. We observe low-dissipation, high-frequency transport up to $\sim\SI{4}{\giga\hertz}$ and $\sim\SI{100}{\micro\meter}$. We fully characterize the plasmon losses and observe in particular a frequency-independent loss mechanism, which we correlate to the residual longitudinal resistance of the edge state and a frequency-dependent dissipation which we ascribe to the presence of charge puddles in vicinity of the edge state.

\paragraph{\label{sec:Device} Sample geometry and DC characterization --}

    \begin{table}[b]
    \caption{\label{tab:Device-list} \textbf{Overview of samples}: Hall resistance $R_{yx}$ and longitudinal linear resistance $r_{xx}$, measured path lengths $L$ for all devices}
    \begin{ruledtabular}
    \begin{tabular}{cccc}
    \textrm{Sample}&
    \textrm{$R_{yx}/R_{K}$} & $r_{xx}$ [\si{\ohm\per\micro\meter}] & $L$ [\si{\micro\meter}]\\
    \colrule
    A & $\mathrm{1.00 \pm 0.03}$ & $\mathrm{2.2 \pm 0.4}$ & 30\\
    B &  $\mathrm{0.97 \pm 0.01}$ & $\mathrm{5.8 \pm 0.4}$ & 50\\
    C &  $\mathrm{1.00 \pm 0.01}$ & $\mathrm{0.2 \pm 0.2}$ & 20, 40, 70\\
    D &  $\mathrm{1.00 \pm 0.01}$ & $\mathrm{1.4 \pm 0.1}$ & 15, 25, 35\\
    \end{tabular}
    \end{ruledtabular}
    \end{table}

    \begin{figure}[t]
        \centering
        \sidesubfloat[\centering]{{\includegraphics[height=3.4cm,keepaspectratio=true]{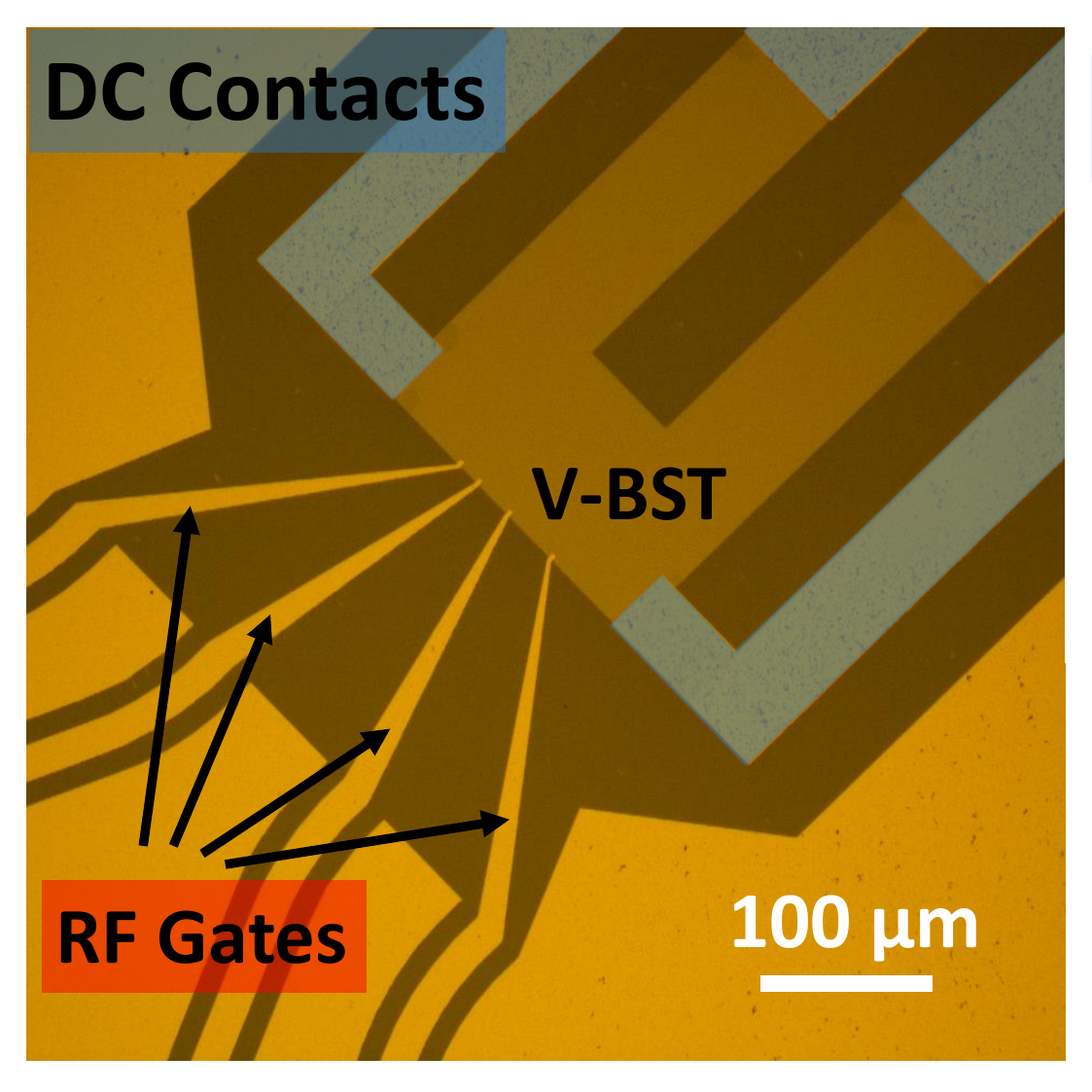}}\,}
        \sidesubfloat[\centering]{{\includegraphics[height=3.4cm,keepaspectratio=true]{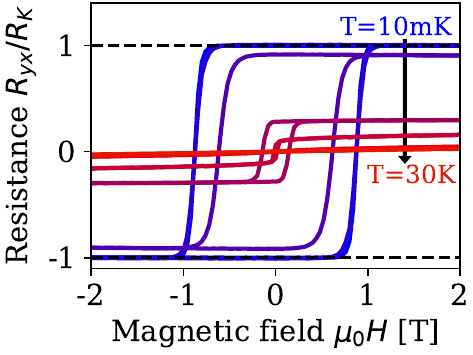}}}\\
        \sidesubfloat[\centering]{{\includegraphics[height=3.4cm,keepaspectratio=true]{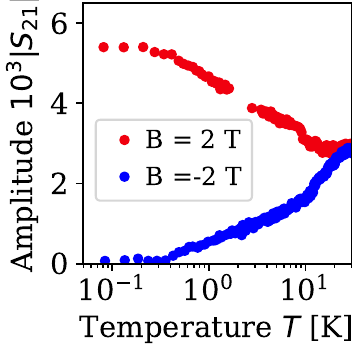}}\,}
        \sidesubfloat[\centering]{{\includegraphics[height=3.4cm,keepaspectratio=true]{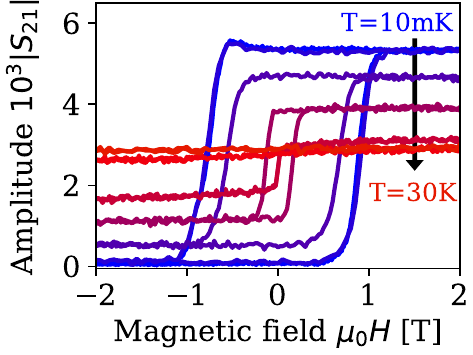}}}
        \captionsetup{justification=justified}
        \caption{\justifying\textbf{Sample geometry, DC and microwave measurements:} a) Optical microscope image of sample D. The $\pi$-shaped mesa is contacted via ohmic contacts highlighted in blue. b) Temperature-dependent field sweeps for the DC transversal resistance $R_{yx}$, which shows quantization until approx. \SI{200}{\milli\kelvin}. The curves correspond to the following temperatures: $T=0.01, 0.3, 1, 3.5, 10, 20$ and \SI{30}{\kelvin}. c) The amplitude for $\mu_0H=\pm\SI{2}{\tesla}$ as a function of temperature $T$ ($L=\SI{20}{\micro\meter}$, $U \simeq\SI{125}{\micro\volt}$, $f\simeq\SI{750}{\mega\hertz}$, device C). d) The amplitude as a function of magnetic field $\mu_0H$ for $L=\SI{20}{\micro\meter}$ and $f=\SI{1}{\giga\hertz}$ on device C for the same temperatures as in b).}
        \label{fig:sample-hysterisis}
    \end{figure}
    
    \begin{figure*}[t]
        \centering
        \sidesubfloat[\centering]{\,{\includegraphics[height=3.6cm,keepaspectratio=true]{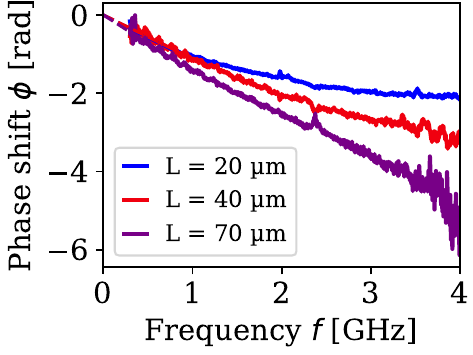}}}\quad
        \sidesubfloat[\centering]{\,{\includegraphics[height=3.6cm,keepaspectratio=true]{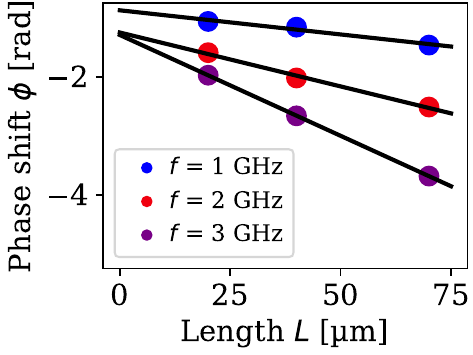}}}\quad
        \sidesubfloat[\centering]{\,{\includegraphics[height=3.6cm,keepaspectratio=true]{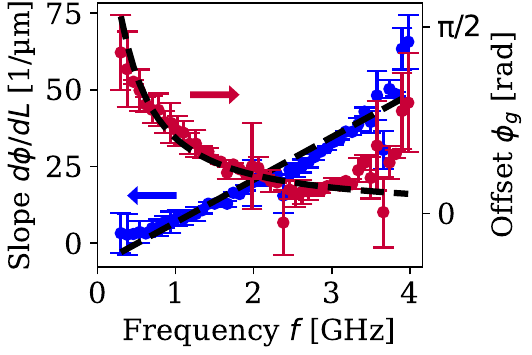}}}

        \caption{\justifying\textbf{Velocity of edge plasmons:} a) phase shift as a function of frequency for a length $L$ of \SI{20}{\micro\meter}, \SI{40}{\micro\meter}, and \SI{70}{\micro\meter} for an excitation voltage of $U \simeq\SI{225}{\micro\volt}$ (device C). The dashed lines show linear fits for frequencies below \SI{0.8}{\giga\hertz}, which we use to subtract the phase at zero frequency. b) Length dependence of phase shift for a frequency of 1, 2, and \SI{3}{\giga\hertz}. The black lines correspond to linear fits. c) Slope $d\phi/dL$ and intercept $\phi_g$ as a function of frequency $f$. The slope is fitted with a  linear function yielding a velocity of $v =\SI{450 \pm 20}{\kilo\meter\per\s}$, while the intercept $\phi_g$ is well fitted by a model of the gate coupling derived from \cite{Gourmelon_2023}, with a gate capacitance $C_g=\SI{19\pm 2}{\femto\farad}$.}
        \label{fig:phase}
    \end{figure*}
    
    The studied devices are fabricated from \SI{8}{\nano\meter}-thick films of V-BST grown on InP (111)A substrates using molecular beam epitaxy (MBE), capped with a \SI{4}{\nano\meter} layer of $\mathrm{Al_2O_3}$. By optimizing the Bi:Sb ratio, the chemical potential is adjusted to reside within the magnetic exchange gap and realize the QAHE without the need for electrostatic gating. After patterning ohmic contacts (\SI{5}{\nano\meter} Pt/\SI{45}{\nano\meter} Au), the V-BST mesa is wet-chemically etched. Capacitively coupled finger gates are realized with an additional \SI{12}{\nano\meter} layer of $\mathrm{Al_2O_3}$, as a gate dielectric and a metallic gate electrode (\SI{5}{\nano\meter} Pt/\SI{45}{\nano\meter} Au). Coplanar waveguides (\SI{5}{\nano\meter} Pt/\SI{45}{\nano\meter} Au) are added to guide microwave signals. Notably, our experimental setup excludes a top gate to prevent screening of edge plasmons, as discussed in \cite{kumada_suppression_2020}. Figure~\ref{fig:sample-hysterisis}a shows a microscope image of one of the devices. Four samples have been investigated with minor variations in the geometry (see Supplementary Material \cite{supplement}), and each of them has been measured in various configurations and path lengths. Thanks to ohmic contacts on the device or on an adjacent Hall bar, the DC longitudinal and transverse (Hall) resistance are determined at a base temperature of $T\simeq\SI{15}{\milli\kelvin}$ before RF measurements (see Table~\ref{tab:Device-list}). All samples exhibit a quantized Hall resistance $R_{yx}=R_K=h/e^2$, with deviations of only a few percent (device B being of lower quality). However, the samples differ more strongly in their residual longitudinal resistance $r_{xx}$ ranging from 0.2 to \SI{5.8}{\ohm\per\micro\meter} in zero field (calculated from the longitudinal $R_{xx}$ measured along a distance of \SI{80}{\micro\meter} between the ohmic contacts). The low values of $r_{xx}$ indicate that charge transport occurs mainly via the edge states and that bulk transport is suppressed. For clarity, we focus on data measured on device C in the rest of the main text. More data can be seen in the Supplementary Material \cite{supplement}, with similar conclusions. The Hall resistance $R_{yx}$ of device C is shown as a function of the out-of-plane magnetic field $\mu_0H$ in Fig.~\ref{fig:sample-hysterisis}b, for various temperatures, revealing a well-known hysteretic pattern \cite{chang_experimental_2013}. In accordance with prior reports \cite{fijalkowski2021a,lippertz-2022}, quantization is visible until onset of bulk transport at approximately $T=\SI{200}{\milli\kelvin}$, while the hysteresis remains visible until the Curie temperature (approximately 20 to \SI{25}{\kelvin} depending on the sample).         
\paragraph{\label{sec:Calib} Microwave setup and calibration --} 
    The microwave measurements are carried out using a vector network analyzer (VNA). The signal at the detection gate is amplified using a cryogenic amplifier (0.3-\SI{14}{\giga\hertz}, +\SI{38}{\decibel}). Non-resonant broadband detection methods require a challenging calibration, which we here briefly summarize (we refer the reader to the Supplementary material for details \cite{supplement}). First, the absence of an isolator (in order to preserve the wide bandwidth of the amplifier) results in standing waves due to the large impedance mismatch between the microwave setup (\SI{50}{\ohm}) and the sample ($R_K\simeq \SI{25.8}{\kilo\ohm}$). Second, a large capacitive stray coupling between the injection and detection ports is observed, which can exceed the transmission via edge plasmons. To correct for stray coupling and other distortions within the setup, we measure reference configurations: 1) a dummy structure without a V-BST mesa serves to calibrate the stray coupling, and 2) a thru-line giving access to the phase shift accumulated along the measurement setup. The calibration of the phase is determined to be robust and device-independent, with a typical accuracy $<0.2$ rad for $f>\SI{1}{\giga\hertz}$. The amplitude of the signal is more device-dependent and suffers from a more inaccurate calibration. We estimate a systematic error of typically \SI{20}{\percent} in the same frequency range (see Supplementary Material\cite{supplement}). In the following, we will always refer to the corrected transmission coefficient $S_{21}$. Due to the asymmetric device geometry, for negative magnetization, plasmons are strongly attenuated along the long edge path ($L>\SI{2}{\milli\meter}$), and we expect to measure only stray signals. We indeed observe, in this configuration, that the signal measured on the device is very close to that of the dummy, validating our experimental approach. For the opposite magnetization direction, we observe a strong additional signal, attributed to edge state transport traveling on shorter distances ($L<\SI{100}{\micro\meter}$). Fig.~\ref{fig:sample-hysterisis}d shows the amplitude of the transmission as a function of the magnetic field $\mu_0H$ for temperature values between \SI{10}{\milli\kelvin} and \SI{30}{\kelvin}, which evidences the hysteretic behavior of the transmission $S_{21}$, showing a high amplitude for positive magnetization ($L=\SI{20}{\micro\meter}$) and no or low signal for negative magnetization ($L>\SI{2}{\milli\meter}$). Fig.~\ref{fig:sample-hysterisis}c shows the temperature dependence of the amplitude at $\mu_0H=\pm\SI{2}{\tesla}$. A constant amplitude is observed upon warming until approximately \SI{200}{\milli\kelvin}. Beyond this temperature range, the amplitude of the edge state signal at $\mu_0H=+\SI{2}{\tesla}$ drops, while a finite and increasing amplitude is measured at $\mu_0H=-\SI{2}{\tesla}$. These two features mark the onset of bulk conductivity as reported in DC transport experiments \cite{Bestwick-2015, lippertz-2022}. Moreover, the coercive field is approximately \SI{0.9}{\tesla} at $T=\SI{10}{\milli\kelvin}$ and decreases with increasing temperature until the hysteresis vanishes at approximately \SI{22}{\kelvin}. These observations in accordance with the DC characterization data shown in Fig.~\ref{fig:sample-hysterisis}b confirm that, after proper calibration, the microwave transmission coefficient associated with transport through the edge state can be singled out.

\paragraph{\label{sec:results} Edge plasmon velocity --}    
    After investigating the amplitude of $S_{21}$, we now briefly analyze its phase to obtain an estimate of the velocity of edge plasmons, before presenting a circuit model describing both in a unified manner. After calibration, the phase shift $\phi=\arg(S_{21})$ depends on the velocity $v$ and the propagation length $L$ of the edge plasmon, as well as an additional contribution $\phi_g$ from the capacitive coupling between finger gate and edge state (see Supplementary Material \cite{supplement}): $\phi(f)=2\pi f L / v+\phi_g(f)$. The phase shift $\phi$, after subtracting the phase at zero frequency, is shown in Fig.~\ref{fig:phase}a for the length of $L=\SI{20}{\micro\meter}$, \SI{40}{\micro\meter}, and \SI{70}{\micro\meter} and for an amplitude $U\simeq\SI{225}{\micro\volt}$. The phase shift increases with increasing frequency $f$ and length $L$, as expected. For each frequency $f$, a linear fit of $\phi$ as a function of $L$ yields the slope $d\phi/dL=2\pi f / v$ and the intercept $\phi_g$ (see example fits in Fig.~\ref{fig:phase}b). A second linear fit of $d\phi/dL$ as function of $f$ (Fig.~\ref{fig:phase}c) yields a velocity of $v =\SI{450 \pm 20}{\kilo\meter\per\s}$. Similarly, the intercept $\phi_g$ is well fitted with a simple model of gate-edge coupling derived by Gourmelon et al. \cite{Gourmelon_2023}, yielding a gate capacitance of $C_g=\SI{19\pm 2}{\femto\farad}$. Despite the uncertainty and inaccuracy in the calibration of multiple samples and configurations, the phase $\phi$ and its dependence on length $L$ and frequency $f$ are well captured by our simple assumptions, giving confidence in our analysis. The estimate of $v$ is comparable to measurements of edge plasmons in Cr-doped BST \cite{martinez_edge_2023,mahoney_zero-field_2017}, and close to the Fermi velocity measured by angle-resolved photoemission spectroscopy in related bulk compounds \cite{jozwiak_spin-polarized_2016,chen_massive_2010}. The velocity $v$ is however lower than the velocity of quantum Hall edge states in ungated devices \cite{kumada2011} where it is increased by electron-electron interaction within the edge state \cite{chamon_sharp_1994,chklovskii_electrostatics_1992}. This could signal the screening of interactions by charge puddles created by fluctuations of the Fermi level, whose role has been exemplified in high-frequency transport of edge states in the quantum Hall and quantum spin Hall insulators \cite{kamata2022, Gourmelon_2023,martinez_edge_2023}. In the next sections, we present an analysis of our results based on a circuit model describing the puddles.

\begin{figure*}[t]
        \centering
        \sidesubfloat[\centering]{{\includegraphics[height=2.8cm,keepaspectratio=true]{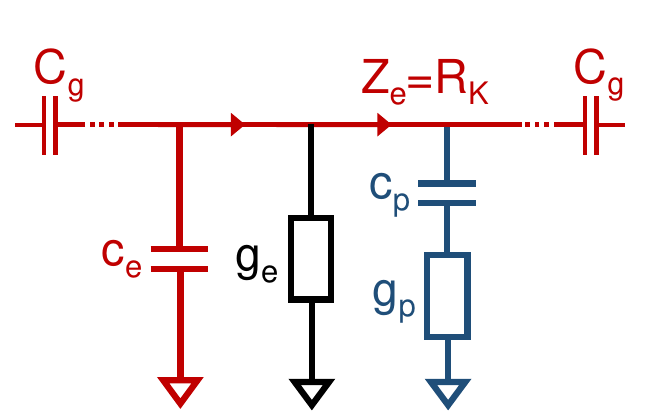}}}\quad
        \sidesubfloat[\centering]{{\includegraphics[height=3.5cm,keepaspectratio=true]{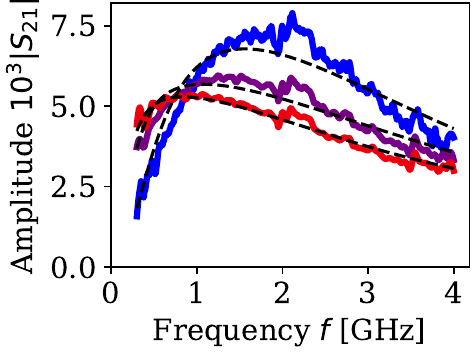} }}\quad
        \sidesubfloat[\centering]{{\includegraphics[height=3.5cm,keepaspectratio=true]{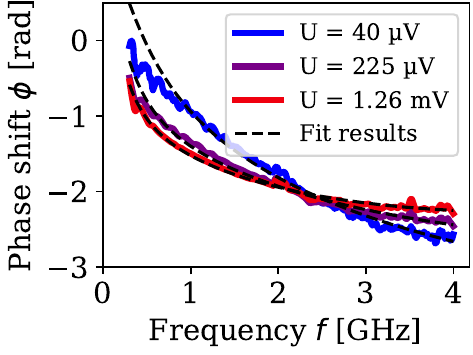} }}
        \caption{\justifying\textbf{Charge puddle model:} a) Circuit diagram, where the edge channel is described by a transmission line of impedance $Z_{e}$ and a line capacitance of $c_{e}$, that is coupled to the setup via a coupling capacitance $C_g$. The interaction with charge puddles is modeled by a conductivity  $g_{e}$ in DC and by a conductivity $g_{p}$ and the line capacitance $c_{p}$ for high frequencies. b)-c) Measurements (solid colored lines) and fits (dashed lines) of the amplitude $|S_{21}|$ and phase $\phi$ (respectively) in device C for $L=\SI{20}{\micro\meter},\, T=\SI{20}{\milli\kelvin}$, and $\mu_0H=\SI{2}{\tesla}$.}
        \label{fig:fit_model}
    \end{figure*}

\paragraph{\label{sec:puddle-model} Charge puddle circuit model --} 
    To treat propagation, dispersion, and dissipation on equal footing, we model our results using the circuit model shown in Fig.~\ref{fig:fit_model}a, adapted from Kumada et al. \cite{kumada2014}. The edge state is modeled as a transmission line of impedance $R_K$ and line capacitance $c_e$, corresponding to the quantum capacitance of the edge state \cite{gourmelon_resonator}. The conductivity $g_{e}$ models frequency-independent dissipation, which we will later confirm to stem from the residual longitudinal resistance seen in DC transport experiments. The conductance $g_{p}$ and the capacitance $c_{p}$ model the interaction with charge puddles as an RC circuit, of characteristic frequency $f_{p}=g_{p}/(2\pi c_p)$, where the total capacitance $c_{p}$ includes both the geometric and quantum capacitances \cite{Dartiailh-2020, Inhofer-2018}. The conductance $g_{p}$ corresponds to the microwave conductance of the bulk. We find the following dispersion relation:
    
    \begin{equation}
        k(f)=   2\pi f R_Kc_e - iR_{K} g_{e}+\frac{2\pi f R_K c_p}{1+\frac{if}{f_p}}
    \end{equation}
    
     With these new circuit elements, the dispersion relation of the edge plasmons is modified in two ways. First, charge puddles slow down edge plasmons, reducing the velocity from $1/R_Kc_e$ to $1/R_K(c_e+c_p)$ (assuming the low-frequency limit $f\ll f_p$). Second, dissipation through puddles takes place, with Im($k(f))\propto f^2$, yielding the same expression as Martinez et al. \cite{martinez_edge_2023}. To reduce the number of fit parameters, we assume that $c_e\ll c_p$ as observed in similar cases \cite{Lin-charge-puddle, Gourmelon_2023}. Additionally, we account for the capacitive coupling between the finger gate and the edge state by the capacitor $C_{g}$, following Ref. \cite{gourmelon_resonator}.

\paragraph{\label{sec:fit} Analysis and fit of the data --} 
    Fig.~\ref{fig:fit_model}b and~\ref{fig:fit_model}c show the amplitude and phase of $S_{21}$ as a function of frequency $f$ for a distance $L=\SI{20}{\micro\meter}$ at different excitation voltages $U$, together with fits using the circuit model. We first discuss the fit for an excitation voltage $U\simeq\SI{225}{\micro\volt}$ (purple curve, same amplitude as in Fig.~\ref{fig:phase}a). The fit shows a good agreement with the data ($R^2=0.98$). It yields a gate coupling capacitance $C_g=\SI{24.4(0.2)}{\femto\farad}$ and a line capacitance $c_p=\SI{102(1)}{\pico\farad\per\meter}$, from which we derive the velocity $v=\SI{380(3)}{\kilo\meter\per\second}$. The velocity $v$ is in agreement with our earlier coarse estimate and with results obtained in Cr-doped BST \cite{mahoney_zero-field_2017,martinez_edge_2023}. The puddles show a characteristic frequency of $f_p=\SI{7.0(0.1)}{\giga\hertz}$, in rough accordance with Ref. \cite{bagchi2019} in which the bulk microwave conductivity of similar compounds has been studied.
    However, the estimate of the fit parameter $g_{e}$, which controls the amplitude of $S_{21}$, is subject to the overall uncertainty on the calibration of the amplitude, as outlined in the supplementary material \cite{supplement}, yielding unreliable absolute values. Hence we will instead focus on relative changes in $g_{e}$. We note that the data for higher excitation voltages $U$ is also very well fitted by our model, though for considerably different parameters. We discuss in the next section the variations of the obtained parameters.
    
\begin{figure}
        \centering
        \includegraphics[width=8.6cm]{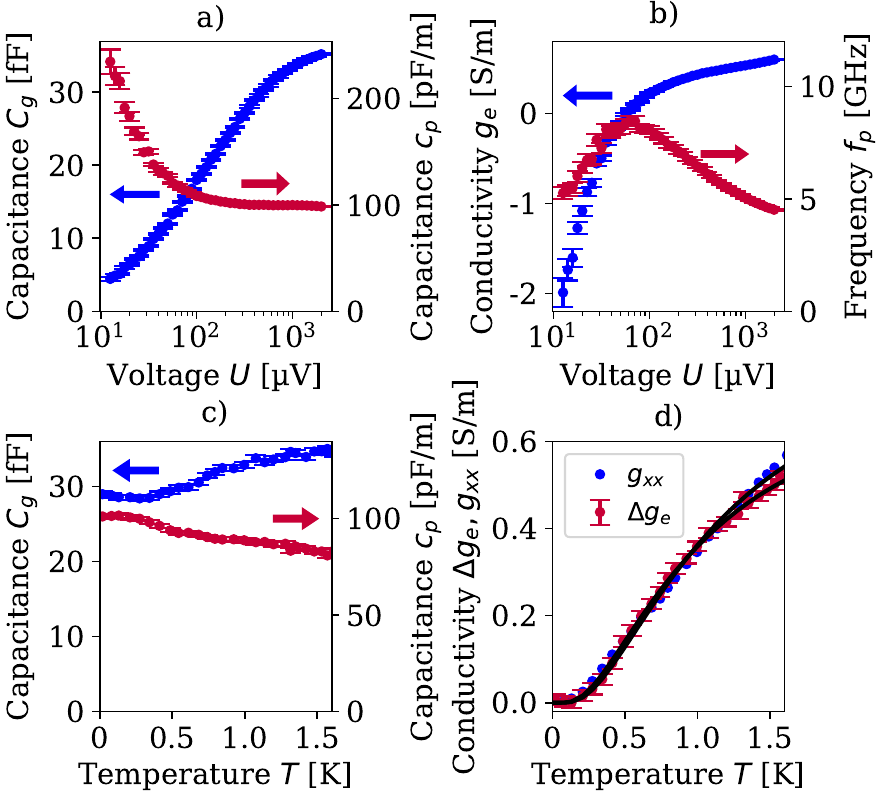}
        \caption{\justifying\textbf{Excitation voltage and temperature dependence of fit parameter:} a)-b) Results of the fit model for the frequency-dependent transmission at different voltages $U$. The coupling capacitance $C_g$ and the conductivity $g_{e}$ increase with an excitation voltage, while the puddle capacitance $c_p$ decreases. All curves show a transition at an excitation voltage of $U\simeq\SI{60}{\micro\volt}$. c) Temperature dependence of the fit parameter $C_g$ and $c_p$ for $U\simeq\SI{125}{\micro\volt}$. d) Temperature dependence of the fitted conductivity $g_{e}$ (red) and of the longitudinal conductance $g_{xx}$ (blue), together with fits based on the variable range hopping model, yielding $T_0=\SI{17.4(0.6)}{\kelvin}$ and $T_0=\SI{15.4(0.2)}{\kelvin}$ respectively.}
        \label{fig:fit}
    \end{figure}

 \paragraph{\label{sec:dissipaton} Dissipation of edge plasmons --} 
    We have previously identified two distinct dissipation mechanisms in the edge state: i) a frequency-independent dissipation described by a series resistance $g_{e}$, ii) a frequency-dependent dissipation due to AC puddle transport, modeled by the parameters $c_p, g_p$ and $f_{p}$. As the temperature or excitation voltage is increased, the dissipation is modified by the increased contribution of bulk carriers due to thermal excitation or breakdown and percolation through the bulk \cite{lippertz-2022}. This statement is substantiated by Fig.~\ref{fig:fit}, which shows the dependence of the fit parameters on the excitation voltage $U$. We observe an enhanced coupling capacitance $C_g$ by nearly one order of magnitude from \SI{4}{\femto\farad} at \SI{10}{\micro\volt} to \SI{35}{\femto\farad} at \SI{2}{\milli\volt}, while $g_{e}$ increases. In the meantime, the puddle capacitance $c_p$ drops from approx. $\SI{235}{\pico\farad\per\meter}$ to approx. \SI{100}{\pico\farad\per\meter}. Both features indicate a more metallic bulk with increased dissipation, due to activation of bulk carriers, and result in an increased velocity by roughly a factor of 2 from \SI{160}{\kilo\meter\per\second} to \SI{390}{\kilo\meter\per\second}. This reduced velocity at low excitation voltage approaches the universal value recently predicted \cite{zhang2024chiral} $v=\frac{1}{2\sqrt{2}R_K \epsilon}\simeq 160-\SI{220}{\kilo\meter\per\second}$ with $\epsilon\simeq 7 \epsilon_0-10\epsilon_0$ the average surrounding dielectric constant. The characteristic puddle frequency $f_p$ shows a maximum of \SI{8.5}{\giga\hertz} at $U\simeq\SI{60}{\micro\volt}$. \\
    We also emphasize that the crossover between these two distinct low and high-voltage regimes occurs for an excitation energy scale $eU\simeq\SI{60}{\micro\electronvolt}$. This agrees well with activation energies found in DC experiments \cite{chang-2015, lippertz-2022, Kawamura-2017, Fox-2018, Bestwick-2015}.
    As seen above, edge state transport is not only broken by an excitation voltage but also by higher temperature $T$, typically at onset of bulk conductivity at approximately $T=\SI{200}{\milli\kelvin}$. This provides means to investigate the relation between the RF bulk conductance $g_e$ and the longitudinal resistance $r_{xx}$ independently measured in DC. To correct for errors in the amplitude calibration, we focus on the relative variation $\Delta g_e(T)=g_e(T)- g_e(T=0)$, and compare it with the line conductance $g_{xx}\simeq r_{xx}/R_{yx}^2$ following the usual relations between conductivity and resistance in a Hall system, in the low-loss limit \cite{yoshioka}. Fig.~\ref{fig:fit}d shows the temperature dependence of $g_e$ and $g_{xx}$, which exhibit similar temperature dependence, namely a strong increase from $g_e, g_{xx}<\SI{0.05}{\siemens\per\meter}$ at low $T$ to larger values $>\SI{0.5}{\siemens\per\micro\meter}$ for $T>\SI{1}{\kelvin}$. The change in $g_e, g_{xx}$ are well fitted by the variable-range-hopping model $g(T)=g_0\exp(-(T_0/T)^{1/2})/T$ describing disordered systems at low temperature, for $T_0\simeq\SI{16}{\kelvin}$. This behavior, observed on all devices, strongly supports a common origin, namely the bulk conductivity as the source of the frequency-independent dissipation. The fit values for $T_0$ and $f_p$ are comparable to those observed in Cr-BST \cite{martinez_edge_2023}, but one order of magnitude lower than those in graphene \cite{kumada2014}, indicating a much stronger disorder. Besides, we also observed an enhanced velocity (low puddle capacitance $c_p$) and coupling for larger temperatures (see Fig.~\ref{fig:fit}c) confirming a metallic behavior at higher temperatures as already observed for large excitation voltages.

\paragraph{\label{sec:discussion} Summary and conclusions --}
    In summary, we reported on the measurement of the characteristics of edge plasmons in the QAH state of V-BST using broadband microwave transport. In a low-voltage and low-temperature regime necessary to observe a robust QAH state, we find a velocity $v\simeq \SI{200}{\kilo\meter\per\second}$, close to the universal value calculated from a disorder-free theory \cite{zhang2024chiral}. This velocity increases as higher excitation voltages or temperatures progressively destroy the QAH state, with values close to those observed in Cr-doped $\mathrm{(Bi_xSb_\text{1-x})2Te_3}$ \cite{mahoney_zero-field_2017,martinez_edge_2023}. Dissipation is mostly dominated by frequency-dependent losses in the low-voltage/low-temperature regime, which we ascribe to charge puddles and model as an RC series circuit. When the QAH state breaks down under increased temperature or excitation voltage, we find that the dissipation rapidly becomes dominated by a frequency-independent mechanism which we relate to the residual longitudinal resistance measured in DC transport experiments.
    This work primarily confirms the propagation of QAH edge plasmons on large distances and high frequencies, here up to \SI{100}{\micro\meter} and \SI{4}{\giga\hertz}, making it a promising platform for plasmonics. It is an important step towards the realization and improvement of QAH-based non-reciprocal components \cite{viola_hall_2014,mahoney_zero-field_2017} and towards the generation of flying Majorana anyons \cite{beenakker_deterministic_2019}.  As a result of the quantitative studies of the devices, we predict that, in the best conditions, circulators and resonators \cite{frigerio2024} based on the measured thin films could reach quality factors approaching 100 at a few tens of \si{\mega\hertz}. Future work will aim to clarify the role of screening gates \cite{kumada_suppression_2020}. The present work also indicates the prominent role of puddles in the dissipation and likely in the breakdown of the quantum anomalous Hall effect. Using microwave signals as probes, future research will investigate the breakdown dynamics. \\

    The supporting data and codes for this article are available from Zenodo \cite{zenodo_data}.

\section{Acknowledgements}
    \begin{acknowledgments}
        The authors warmly thank C. Dickel, J. Hemberger, M. Hashisaka, N. Kumada, H. Kamata, G. Fève, E. Frigerio, and G. Ménard for insightful discussions.
        This work has been supported by the ERC, under contract ERC-2017-StG "CASTLES", Germany’s Excellence Strategy (Cluster of Excellence Matter and Light for Quantum Computing ML4Q, EXC 2004/1 - 390534769), the DFG (SFB1238 Control and Dynamics of Quantum Materials, 277146847, project A04 and B07).
    \end{acknowledgments}
\bibliography{Literature}
\clearpage
\pagebreak

\appendix
\renewcommand{\thefigure}{S\arabic{figure}}
\renewcommand{\thetable}{S\arabic{table}}
\setcounter{figure}{0}
\setcounter{table}{0}

\section{Supplementary Material}\label{sec:appendix}
    
\subsection{Sample preparation and DC properties}
    
    The device fabrication involves three steps: deposition of ohmic contacts, etching of V-BST, and deposition of gates. The ohmic contacts are patterned using electron-beam lithography, followed by sputtering \SI{5}{\nano\meter} of Pt and \SI{45}{\nano\meter} of Au. To structure the V-BST film, we perform optical lithography followed by wet-chemical etching in Transene Aluminum Etchant type-D to etch the $\mathrm{Al_2O_3}$ and a Piranha acid solution to etch the V-BST. Subsequently, the entire device is capped with an additional \SI{12}{\nano\meter} layer of $\mathrm{Al_2O_3}$, serving as a gate dielectric for capacitive contacts and as a protective layer for regions that were uncovered during fabrication. 
    The samples have two distinct designs. Samples A and B feature a $\pi$-shaped Hall bar with four ohmic contacts and two capacitive contacts. The ohmic contacts facilitate DC characterization, while the \SI{3}{\micro\meter}-wide capacitive contacts are used for microwave experiments. On the other hand, Samples C and D are equipped with a total of four capacitive contacts each. Additionally, these samples include Hall bars with ohmic contacts for pure DC characterization. They have a width of \SI{50}{\micro\meter} and we measure the longitudinal resistance on a length of \SI{80}{\micro\meter}. The longitudinal resistance $R_{xx}$ and the transversal (Hall) resistance $R_{yx}$ are shown in Fig.~\ref{fig:DC-hysterisis}, after (anti)-symmetrization. To compare the longitudinal resistance $R_{xx}$ to the conductivity $g_e$, we compute the residual bulk conductivity $g_{xx}$:
    
    \begin{equation}
        g_{xx}=\frac{1}{L}\frac{R_{xx}}{R_{xx}^2+R_{yx}^2}\simeq\frac{r_{xx}}{R_{yx}}
    \end{equation}
     We find good agreement between $g_{xx}$ and $g_e$ at low temperatures, where $R_{xx}$ is small compared to $R_{yx}$.
     
\subsection{RF Setup and Calibration}
\begin{figure}[t]
        \centering
        \includegraphics[width=8.0cm]{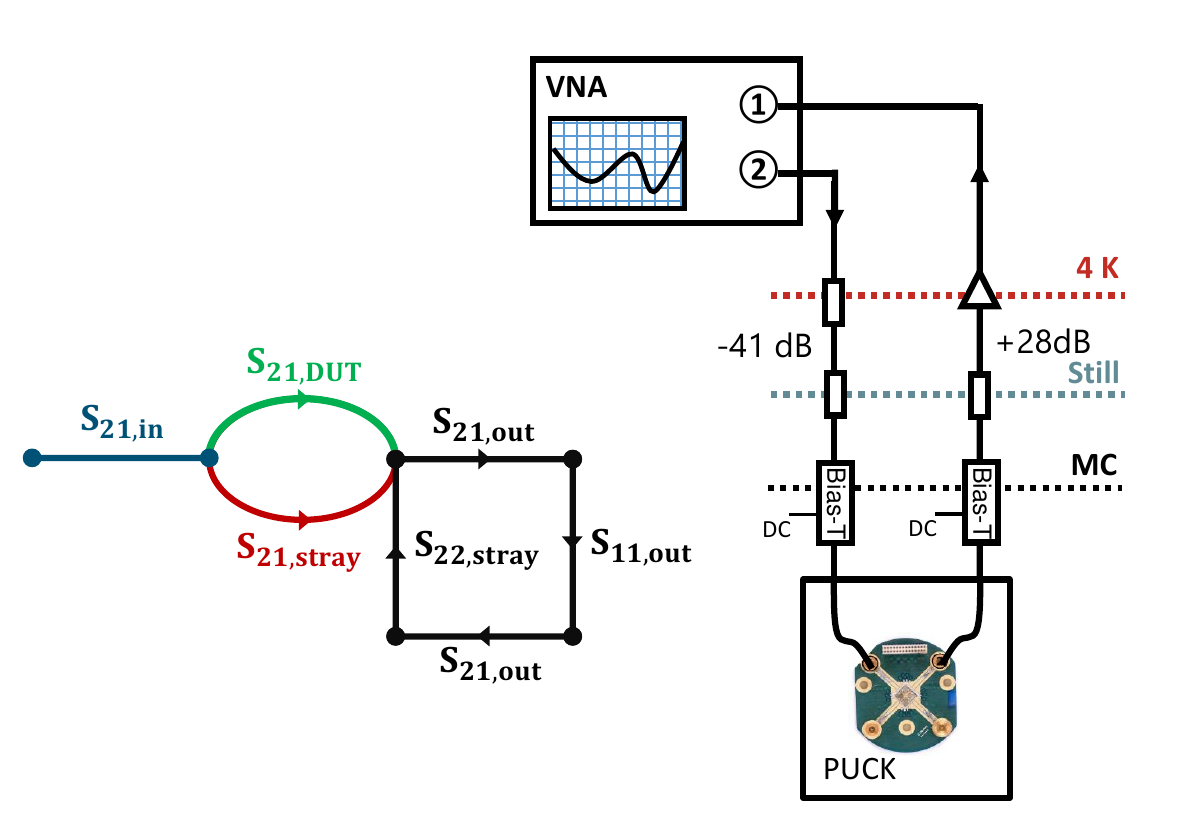}
        \caption{\justifying \textbf{Microwave setup:} Right: Microwave setup in the dilution refrigerator. The incoming signal is attenuated at all temperature stages. The outgoing signal is attenuated by \SI{10}{dB} before being amplified by a cryogenic low-noise amplifier with a bandwidth of 0.3 to \SI{14}{\giga\hertz}. The sample is glued on a printed circuit board and connected via wire bonds. Left: Simplified signal flow graph, that is used for correcting the signal}
        \label{fig:Rf-setup}
    \end{figure}

        The whole RF setup is shown in Fig.~\ref{fig:Rf-setup}. The sample is placed in a printed circuit board with coplanar waveguides. It is contacted via wire bonds and measured in a dilution refrigerator with a base temperature of approx. \SI{10}{\milli\kelvin} and connected via filtered DC lines and broadband coaxial cables. The in-going signal is attenuated at all temperature stages. The outgoing signal is amplified at the \SI{4}{\kelvin} stage ({\it LNF-LNC0.3\_14B} from Low Noise Factory). The scattering parameters are measured using a vector network analyzer (Keysight Streamline P9375A) at room temperature. We estimate that the microwave power $P$ is attenuated by approx. \SI{45}{dB} (\SI{41}{dB} by the attenuators and an additional \SI{4}{dB} by the cables). We note that to preserve the largest bandwidth, no isolator or circulator has been placed in the readout port. In this work, we use the applied voltage on the samples, which is given by $U=\sqrt{10^{-45/10} \cdot Z_g \cdot P }$. We identify three main distortions of an applied signal in our setup:

    \begin{figure}[t]
        \centering
        \includegraphics[height=4cm]{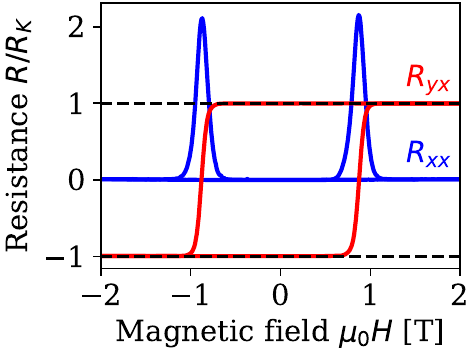}
        \caption{\justifying \textbf{DC Hall Measurement:} The longitudinal resistance $R_{xx}$ (blue) and the transversal (Hall) resistance $R_{yx}$ (red) are shown as a function of the applied magnetic field $\mu_0H$ for sample C. The curves were obtained with a lock-in amplifier with a current amplitude $I=\SI{5}{\nano\ampere}$ and a frequency $f=\SI{7}{\hertz}$. }
        \label{fig:DC-hysterisis}
    \end{figure}

\begin{figure}[t]
        \centering
        \includegraphics[width=6.0cm]{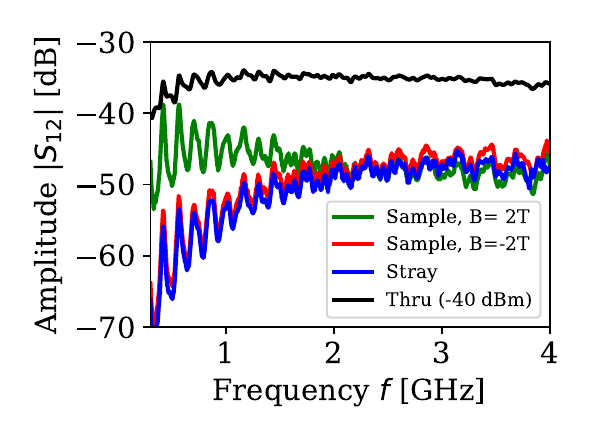}
        \caption{\justifying \textbf{Calibration measurements:} The raw data of the scattering coefficients for the sample at $\mu_0H=\pm\SI{2}{\tesla}$ and for the reference devices are shown as colored solid lines. The through trace is offset by \SI{-40}{\decibel} for clarity.}
        \label{fig:Rf-Calib}
    \end{figure}

    \begin{itemize}
        \item \:The damping and phase-shift of all microwave components (coaxial cables, attenuators, bias-tees, amplifier).
        \item \:Standing waves between the sample and amplifier. They are visible as oscillations in the VNA spectra as shown in Fig.~\ref{fig:Rf-Calib}. This is due to the large impedance mismatch between edge state ($Z_e\simeq\SI{25.8}{\kilo\ohm}$) and coplanar waveguide ($Z_g=\SI{50}{\ohm}$) and the relatively large reflection $S_{11}>0.1$ of the amplifier. To dampen the standing wave, We add a \SI{10}{dB} attenuator in front of the amplifier.
        \item \:Stray coupling between signal lines on the chip. In the microwave regime, the coplanar waveguides on the chip will act as antennas and be directly coupled to each other. This effect is increasing with frequency.
    \end{itemize}

    To correct for the distortions mentioned above, we model our setup using a signal graph as shown in Fig.~\ref{fig:Rf-setup}. We model the transmission from the input and output lines via $S_{21,in}$ and $S_{21,out}$. The stray coupling is described via $S_{21,stray}$, and the sample (edge state and capacitive coupling to the gate) is described by $S_{21,DUT}$. To account for the standing waves, we added a loop on the output side with the reflection coefficients $S_{11,out}$ and $S_{22,stray}$. The whole transmission is given by:
    
    \begin{equation}
        S_{21}=\frac{S_{21,in}\cdot(S_{21,DUT}+S_{21,stray})\cdot S_{21,out}}{1-S_{21,out}^2 \cdot S_{11,out} \cdot S_{22,stray}}
    \end{equation}
    
    Though it simplifies the real setup, this approach allows us to correct the signal by measuring two reference samples. The first reference sample is a through-dummy. Here the injection and detection ports are directly connected via a coplanar waveguide, thus yielding the response of the whole setup without sample or stray coupling:
    
    \begin{equation}
        S_{21,thru}=S_{21,in}\cdot S_{21,out}
    \end{equation}
    The second reference is a stray dummy, a sample identical to those measured, but for which the V-BST is entirely etched away. This reference shows the stray coupling and the standing waves:
    \begin{equation}
        S_{21,dummy}=\frac{S_{21,in}\cdot S_{21,stray}\cdot S_{21,out}}{1-S_{21,out}^2 \cdot S_{11,out} \cdot S_{22,stray}}
    \end{equation}
    
    \begin{figure*}[ht]
        \centering
        \sidesubfloat[\centering]{{\includegraphics[height=3.8cm]{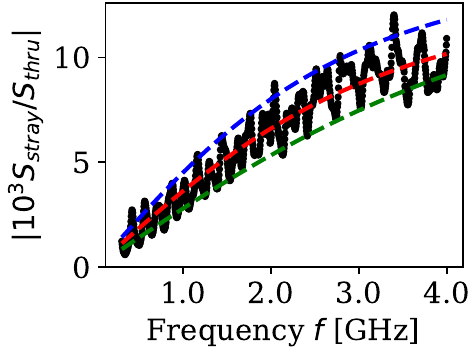}}}
        \sidesubfloat[\centering]{{\includegraphics[height=3.8cm]{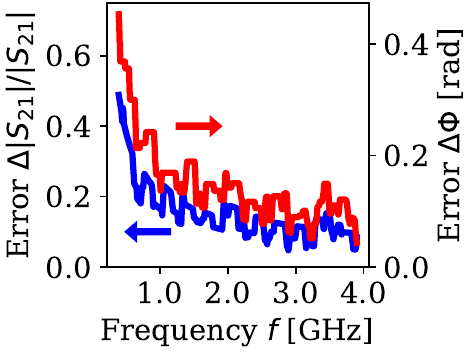}}}
        \sidesubfloat[\centering]{{\includegraphics[height=3.8cm]{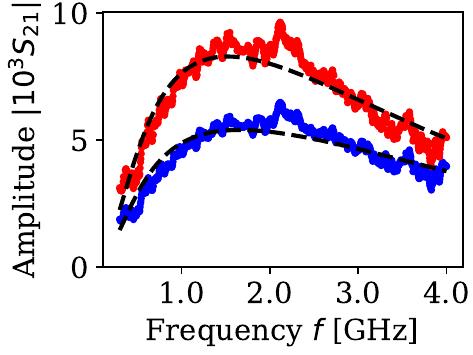}}}
        \caption{\justifying\textbf{Fit of capacitive coupling:} a) The corrected stray signal $S_{stray}/S_{thru}$ is shown as black dots. The red line shows the fit with the stray function corresponding to a stray capacitance of \SI{135}{\femto\farad}. To estimate the error, we performed another fit with the upper and lower bounds of the stray signal, which are shown in blue and green respectively. b) Standard deviation of the amplitude and phase calculated from the lower and upper bound of the corrected stray signal. c) Amplitude as a function of the frequency for the correction using the lower and upper bound of the stray signal. The dashed lines show the fit using the circuit model that we use to estimate the lower and upper bound of the fit parameter.}
        \label{fig:error_correction}
    \end{figure*}
    
    \begin{figure}[b]
        \centering
        \includegraphics[width=5.0cm]{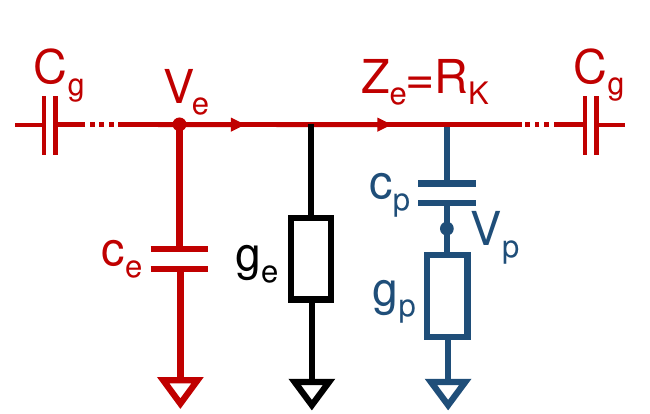}
        \caption{\justifying\textbf{Circuit model:} Same circuit model as in Fig.~\ref{fig:fit_model}. Additionally, the knots where we calculate current conservation are shown. They are at a voltage $V_e$ and $V_p$ for the edge state and the puddles respectively. }
        \label{fig:circuit-model}
    \end{figure}
    The VNA spectra of the sample at both magnetization directions, the through dummy, and the stray dummy are shown in Fig.~\ref{fig:Rf-Calib}. The stray dummy and the sample at negative magnetization ($\mu_0H=\SI{-2}{\tesla}$) show nearly the same transmission, confirming that we can also use the measurement with the long path as a reference.
    The stray coupling and the standing waves need to be disentangled to correct the signal. To do so, we fit the amplitude $|S_{21,dummy}/S_{21,thru}|$ to the scattering parameter of a capacitor $C$ in a \SI{50}{\ohm}-network with an amplitude $A$, which is given by:
    
    \begin{equation}
        S_{21,capa}=\frac{A}{1+\frac{0.01j}{2\pi\omega C}}
    \end{equation}
    Using the fit result $S_{21,capa}$, we obtain the scattering parameter of the standing waves $S_{21,osci}$:
    \begin{equation}
        S_{21,osci}=\frac{S_{21,stray}}{S_{21,thru}\cdot S_{21,capa}}
    \end{equation}
    And finally, the corrected transmission of the actual device reads:
    \begin{equation}
        S_{21,DUT}'=\frac{(S_{21,DUT}-S_{21,stray})\cdot S_{21,capa}}{S_{21,stray}}
    \end{equation}
    
    The main limitation of this correction lies in the fits of the stray coupling, which underlies a significant error. Fig.~\ref{fig:error_correction}a shows the stray signal and the fit result. Due to the standing waves, there is a large discrepancy between the lower and upper bounds of the signal. To estimate the error of the correction, we estimate the lower and upper bounds of the amplitude and the phase. The standard deviation calculated from these bounds is shown in Fig~\ref{fig:error_correction}b. We correct the measured dispersion with both the lower and upper bound and fit it with the charge puddle model to estimate the introduced error. For an excitation voltage of $U\simeq\SI{40}{\micro\volt}$, we find the following lower and upper bounds: $g_e$~=~$[0.27\pm0.02,-0.67\pm0.02]$ \si{\siemens\per\meter}, $C_g$~=~$[10.8\pm0.2,10.5\pm0.2]$ \si{\femto\farad}, $c_p$~=~$[126\pm2,136\pm2]$ \si{\pico\farad\per\meter}, and $f_p$~=~$[10.1\pm0.4,7.3\pm0.2]$ \si{\giga\hertz}. Especially the parameter $g_e$ changes drastically, as it models the frequency-independent dissipation. \\
    Additionally, the calibration also changes between subsequent cooldowns, introducing a small systematic error, that we cannot reliably evaluate. However, this gives only an offset on amplitude and phase and does not contradict our results.
    
     It is also worth mentioning, that our procedure does not correct for the capacitive coupling between the coplanar waveguide and the edge state via the finger gate, which must therefore be taken into account as part of the response of the device.\\
    
\subsection{Derivation of circuit model}
    To find the dispersion relation of our distributed circuit model (Fig.~\ref{fig:circuit-model}), we use Kirchoff's current conservation law, to find the chiral wave equation:
    \begin{equation}
            \begin{split}
            c_e \partial_t V_e+\frac{1}{R_K}\partial_x V_e+g_eV_e+c_p(\partial_t V_e-\partial_t V_p)=0 \\
            g_p V_e + c_p(\partial_t V_p-\partial_t V_e)=0 
            \end{split}
    \end{equation}
    The chirality of the edge channel is imposed in the first equation. Searching for plane wave solutions  $V_{e/p}=V_{e/p,0} \cdot e^{i(\omega t - k x)}$, we rewrite the chiral wave equation:
    \begin{equation}
            \begin{split}
            i\omega c_e V_e-\frac{ikV_e}{R_K}+g_eV_e+i\omega c_p(V_e-V_p)=0 \\
            g_p V_e  +i\omega c_p(V_p - V_e)=0 
            \end{split}
    \end{equation}
    Solving the equation yields the dispersion relation:
    \begin{equation}
        \label{eq:dispersion}
        k(\omega)=R_Kc_e\omega-iR_K g_{e}+\frac{ R_K c_p \omega}{1+\frac{i\omega c_p}{g_p}}
    \end{equation}
    The dispersion relation comprises three different terms. The first term describes the linear dispersion of the bare edge state, in absence of dissipation and puddles. The second term governs the frequency-independent dissipation due to $g_e$. The last term features the interaction with charge puddles with a conductivity $g_p$ and a capacitance $c_p$. We introduce the characteristic frequency $f_p=c_p/(2\pi g_p)$.

\subsection{Fitting of data}
    
    \begin{figure*}[ht]
        \centering
        \includegraphics[width=12.0cm]{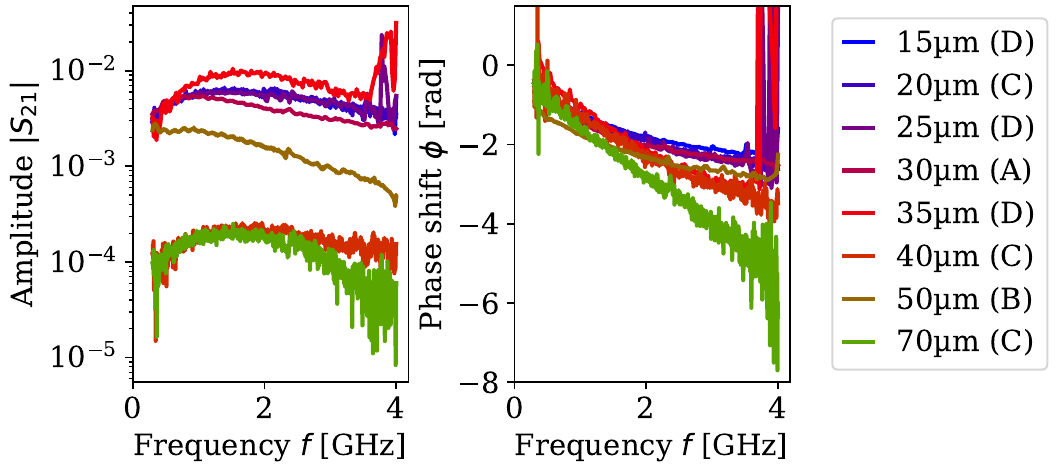}
        \caption{\justifying\textbf{Amplitude and phase vs. frequency for all samples:} Corrected amplitude (left) and phase (right) as a function of the frequency for all measured devices. The curves are obtained at base temperature and an excitation voltage of $U\simeq\SI{125}{\micro\volt}$.}
        \label{fig:amplitude-frequency}
    \end{figure*}   
    As mentioned earlier, our calibration does not correct for the capacitive coupling between the edge state and the capacitive contacts. To account for this in our model, we are calculating the scattering parameter of the capacitive coupling as derived by Gourmelon et al. 
    \cite{gourmelon_resonator}. They describe the coupling by a series of a geometric capacitance $C_{geo}$ and a quantum capacitance $C_e$. To reduce the number of fit parameters, we instead use only the effective coupling capacitance $C_g$, which is given by $C_{geo}\cdot C_{e}/(C_{geo}+C_{e})$. The scattering parameters between the gate and the edge state are:
    \begin{equation}
            S_{eg} = \frac{ 2i \omega Z_g C_g / \Lambda_e}{1 + i\omega C_g Z_g / \Lambda_e} = \frac{Z_g}{R_K} S_{ge}
    \end{equation}
    with $\Lambda_e=1+i\omega R_K C_g$ and $Z_g$=\SI{50}{\ohm}, with a single fit parameter $C_g$. For large frequencies $\omega \gg 1/Z_gC_g$, the coupling is constant and the scattering parameter saturates at $S_{ge}\approx 2$ and $S_{ge}\approx 2Z_g/Z_0$ due to the impedance mismatch between the gate electrode and the edge channel. The attenuation and the phase shift in the edge state is described by its dispersion and is defined as:
    \begin{equation}
        S_{ee}=e^{-ik\cdot L}
    \end{equation}
    where $k(\omega)$ is given by the dispersion in equation~(\ref{eq:dispersion}) and $L$ by the path length. In our calibration, the phase measured in the stray-coupling configuration is used as a phase reference, therefore introducing a phase offset of $\pi/2$ due to the capacitive coupling. To account for that, we write the following total scattering parameter $S_{21}$ as:
    \begin{equation}
        S_{21}= S_{eg} \cdot S_{ee} \cdot S_{ge} \cdot e^{-i\pi/2} \
    \end{equation}
    We fit the complex scattering parameter $S_{21}$ with the calibrated complex signal $S_{21,DUT}'$ using the \textit{curve\_fit} function of the \textit{scipy} python package. 

\subsection{Other samples and reproducibility}
\begin{figure}[t]
        \centering
        \includegraphics[width=8cm]{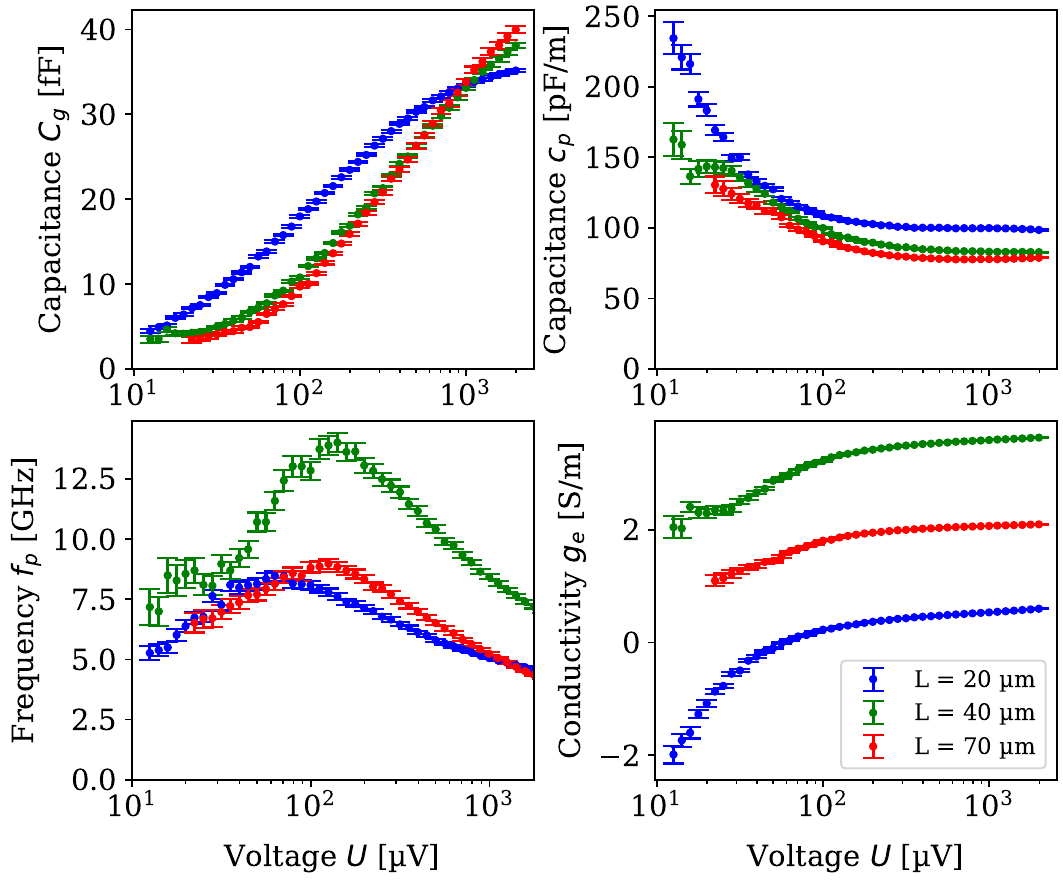}
        \caption{\justifying\textbf{Excitation voltage dependence for Sample C:} The fit parameter $C_g$, $c_p$, $f_p$, and $g_e$ are shown as a function of the excitation voltage $U$ for a path length of $L$~=~\SI{20}{\micro\meter}, \SI{40}{\micro\meter}, and \SI{70}{\micro\meter} on Device C. All curves were obtained at base temperature. The different path lengths show qualitatively the same behavior. The conductivity $g_e$ shows a large offset between different samples due to improper calibration.}
        \label{fig:power-Dec4a}
    \end{figure}

\begin{figure}[b]
        \centering
        \includegraphics[width=8cm]{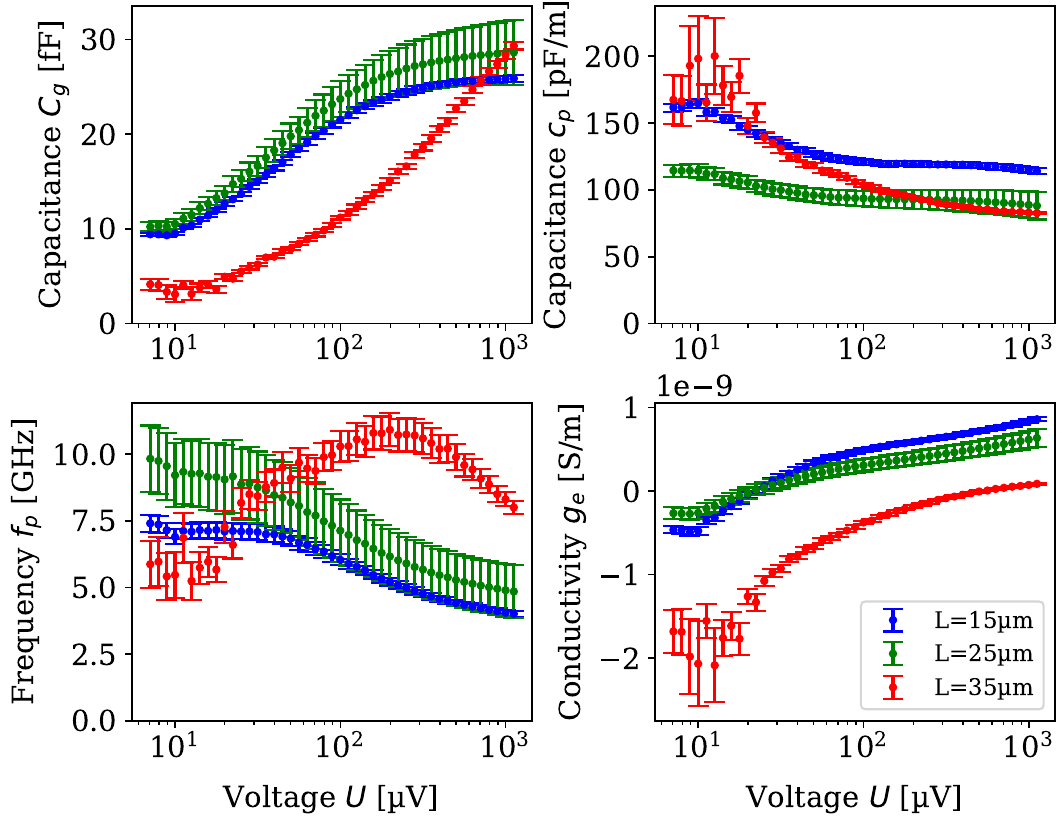}
        \caption{\justifying\textbf{Excitation voltage dependence for Sample D:} The fit parameter $C_g$, $c_p$, $f_p$, and $g_e$ are shown as a function of the excitation voltage $U$ for a path length of $L$~=~\SI{15}{\micro\meter}, \SI{25}{\micro\meter}, and \SI{35}{\micro\meter} on Device D. All curves were obtained at base temperature. The different path lengths show qualitatively the same behavior. The measurement for $L=\SI{35}{\micro\meter}$ shows an offset with respect to the x-axis, which is likely due to an additional attenuation of the in-going signal that is not captured in our calibration measurement.}
        \label{fig:power-Feb6a}
    \end{figure}

\begin{figure}[t]
        \centering
        \includegraphics[width=8cm]{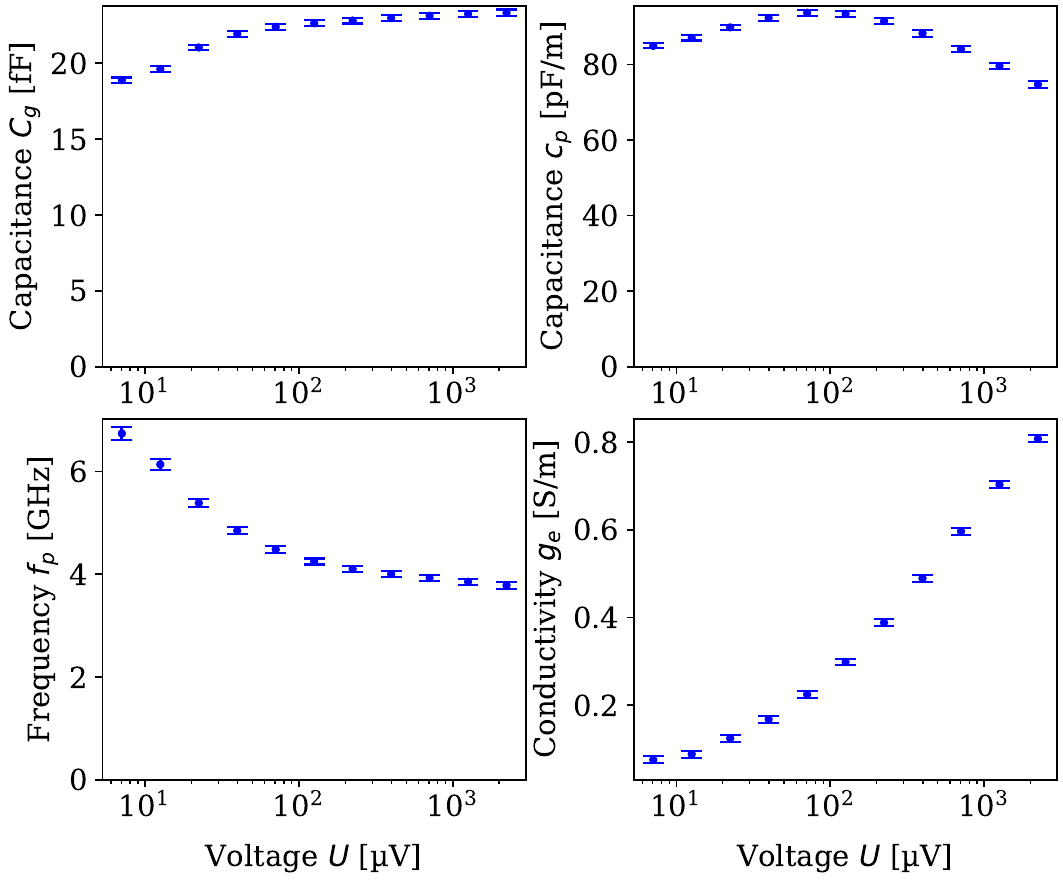}
        \caption{\justifying\textbf{Excitation voltage dependence for Sample A:} The fit parameter $C_g$, $c_p$, $f_p$, and $g_e$ are shown as a function of the excitation voltage $U$ for a path length of $L$~=~\SI{30}{\micro\meter} for Sample A. All curves were obtained at base temperature.}
        \label{fig:power-Apr4a}
    \end{figure}

\begin{figure}[b]
        \centering
        \includegraphics[width=8cm]{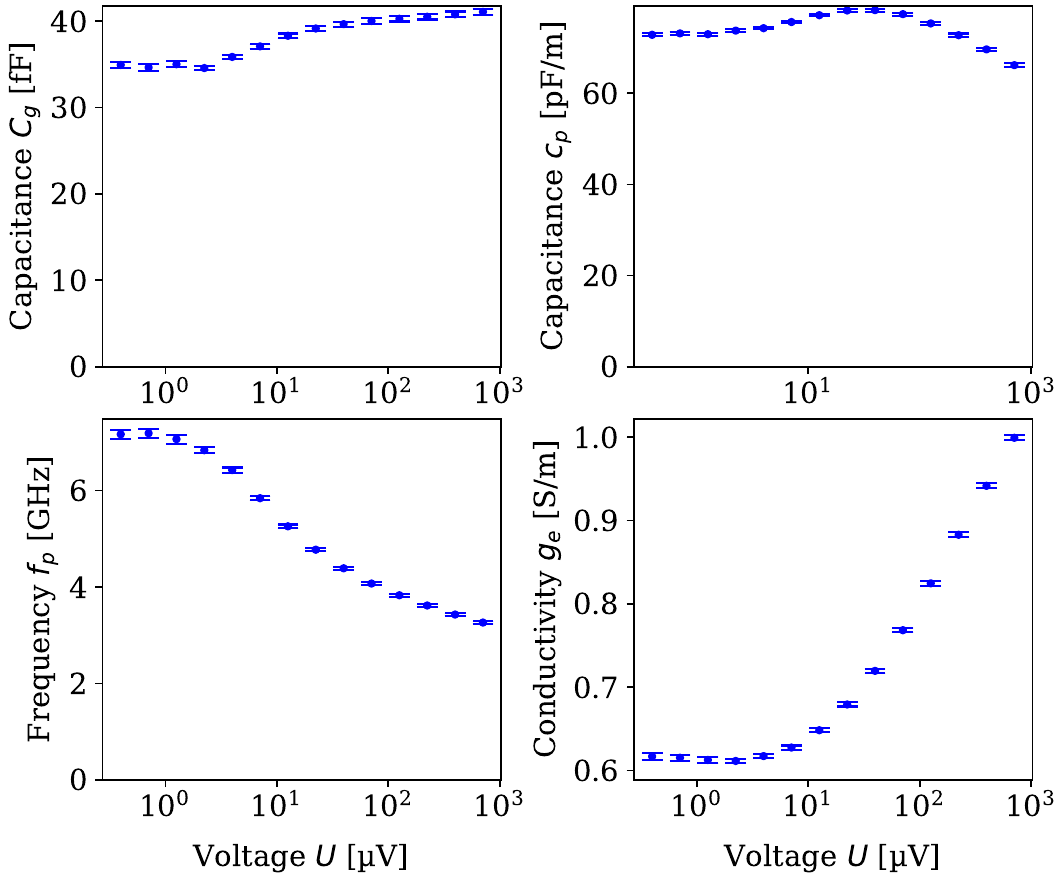}
        \caption{\justifying\textbf{Excitation voltage dependence for Sample B:} The fit parameter $C_g$, $c_p$, $f_p$, and $g_e$ are shown as a function of the excitation voltage $U$ for a path length of $L$~=~\SI{50}{\micro\meter} for Sample B. All curves were obtained at base temperature.}
        \label{fig:power-Apr5b}
    \end{figure}

\begin{figure}[t]
        \centering
        \includegraphics[width=8cm]{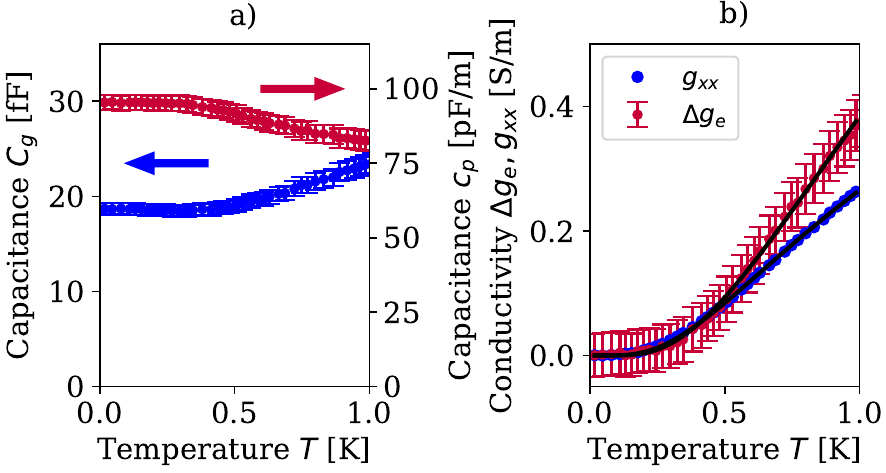}
        \caption{\justifying\textbf{Temperature dependence for Sample D:} a) The fit parameter $C_g$ and $c_p$ are shown as a function of the temperature for an excitation voltage of $U\simeq\SI{125}{\micro\volt}$. The data was acquired for a path length of $L=\SI{25}{\micro\meter}$ on sample D. b) Temperature dependence of $g_e$ is shown in red and the measured longitudinal conductivity $g_{xx}$ is shown in blue. The fit parameter and the measured curve show the same temperature dependence and are only separated by a small margin. The black lines show fits based on the variable range hopping model, yielding $T_0=\SI{25.9(0.4)}{\kelvin}$ and $T_0=\SI{17.7(0.2)}{\kelvin}$ for $g_e$ and $g_{xx}$ respectively.}
        \label{fig:Temperature-Feb6a}
    \end{figure}

    In this section, we discuss the results of the other samples and demonstrate the reproducibility of our measurements. The amplitude and phase for all devices and lengths are shown in Figure~\ref{fig:amplitude-frequency}, for the same applied voltage of approx. \SI{125}{\micro\volt}. The amplitude and phase are device-dependent and suffer from insufficient calibration. Hence we focus on the length dependence within one device. However, all devices and distances show a similar frequency dependence: A maximum amplitude around \SI{1}{\giga\hertz} and a decreasing phase with increasing frequency. \\
    We analyze the excitation voltage dependence for all devices by fitting the scattering parameters to the circuit model. The resulting parameters are shown in Figures \ref{fig:power-Dec4a}, \ref{fig:power-Feb6a}, \ref{fig:power-Apr4a}, and \ref{fig:power-Apr5b}. We observe an enhanced coupling capacitance for all samples with increasing excitation voltage $U$. Samples C and D, which have the lowest residual resistance $r_{xx}$, show the most pronounced effects. We attribute this to samples A and B already being in the breakdown regime at low excitation voltages. Similarly, the charge puddle capacitance $c_{p}$ for samples C and D decreases at low excitation voltage and remains nearly constant above $U\simeq\SI{100}{\micro\volt}$, corresponding to a velocity of approximately \SI{390}{\kilo\meter\per\second}. For samples A and B, the capacitance $c_{p}$ is nearly constant at approximately 70-\SI{90}{\pico\farad\per\meter}, confirming that they are already in the breakdown regime.
    The characteristic frequency $f_p$ ranges between approximately \SI{4}{\giga\hertz} and \SI{15}{\giga\hertz} depending on the sample, excitation voltage, and length. For the samples with low residual resistance $r_{xx}$, we observe the highest characteristic frequency $f_p$ in the voltage regime attributed to the breakdown regime.
    Furthermore, we observe an increase in DC dissipation, described by the conductivity $g_{e}$, with increasing excitation voltage. Notably, a plateau for very low excitation voltages is observed in sample B. Our studies show a clear sample dependence, attributed to the different residual resistances $r_{xx}$. This emphasizes the importance of minimizing $r_{xx}$ to realize high-quality devices with low dissipation. However, the study also indicates that the plasmon velocity reaches approximately \SI{400}{\kilo\meter\per\second} in the breakdown regime for all devices. Also, the findings from other studies confirm this velocity, indicating that they are in the breakdown regime \cite{mahoney_zero-field_2017,martinez_edge_2023}.
    Temperature dependence measured for sample D, as shown in Figure \ref{fig:Temperature-Feb6a}, reveals enhanced coupling and decreased charge puddle capacitance with increasing temperature. The temperature dependence of the fit parameter $r_{e}$ and the measured resistance $r_{xx}$ exhibit a constant and low resistance at low temperatures and onset of bulk conductivity at approximately \SI{200}{\milli\kelvin}, confirming the validity of the circuit model.

    \clearpage


    \clearpage

\end{document}